*Costantino Sigismondi*

# Umbra in partial lunar eclipses at moonrise

Costantino Sigismondi *(ICRA/Sapienza/Liceo G. Ferraris, Roma)*

**Abstract** The darkness of the umbra of the lunar eclipse of August 7 2017 corresponded to the predictions of A. Danjon for the Sun at its minimum. It appeared partial at moonrise in Rome similarly to the one of April 3, 33 AD in Jerusalem, supposed following the Crucifixion of Christ. The hypothesis of Sun at maximum activity for that historical eclipse is discussed.

**Sommario** L'ombra dell'eclissi parziale del 7 agosto 2017 era molto scura, compatibile con la scala di Danjon corrispondente al minimo solare. Avvenuta con la Luna al suo sorgere a Roma, è stata simile a quella del 3 aprile 33 d.C. a Gerusalemme, che daterebbe la Crocifissione di Cristo. L'ipotesi del Sole al massimo di attività per quella storica eclissi è discussa.

## Introduction: a partial eclipse at moonrise

Humphreys and Waddington in 1983[1] proposed to interprete the account of blood-stained Moon (Acts, 2:20) described by St. Peter after the Crucifixion of Christ, as the effect of the partial phase of the eclipse at moonrise in Jerusalem on friday 3 april 33 AD. Schaefer (1990)[2] replied that the reddish Moon was due to the intense atmospheric extinction of 40 equivalent airmasses. Sigismondi (2012)[3] with measurements in Jerusalem from Temple Institute, disagreed this high airmass evaluation.

The partial lunar eclipse of august 7, 2017 at moonrise in Rome has given the occasion for the comparison with the 33 AD one.

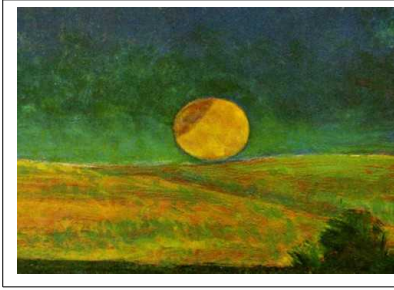

The Moon appearing to "turn to blood" at the lunar eclipse on Friday, 3 April A.D. 33, viewed from Jerusalem after sunset, at the rising of the full moon of Passover. Atmospheric refraction shapes the Moon elliptical. Paint in JohnPratt.com and Nature **306**

---

## The Danjon scale[4] of the lunar eclipses

At solar maximum the lunar eclipses are less dark than at minimum activity, and in 2017 the Sun is at minimum, with a very dark umbra observed in the partial phase of August 7. The eclipse of 33 AD in order to be blood-stained was with Sun at maximum.

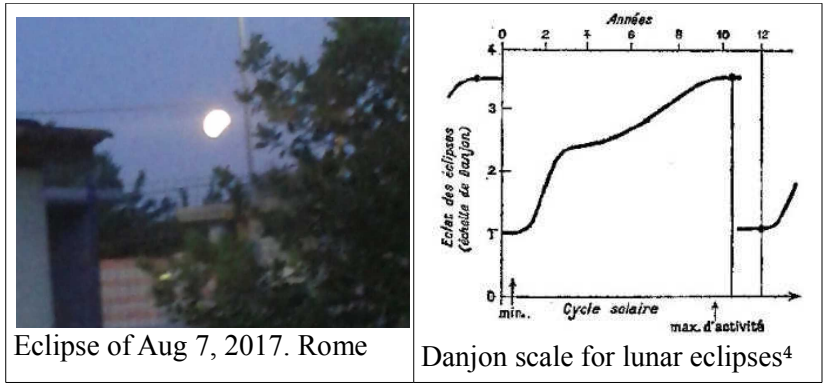

Eclipse of Aug 7, 2017. Rome | Danjon scale for lunar eclipses[4]

## Solar cycle stability along millenia

The 11 year solar cycle is considered stable. Even after Maunder minimum it reappeared synchronously. It it probably linked to the inner[5] and outer rotation of the Sun.

The modulations on the solar cycle have to be matched with the ones of the planets.[6] The phase of a solar maximum in 33AD, deduced from the blood-stained eclipsed Moon could fix the solar cycle phase two millenia ago or test the model of the sunspot-planets influence on the cycle's irregularities.[7]

**Acknowledgments:** to Giuseppina de Felice Proia for the photo.
**References:** C. Sigismondi, Astronomia nei Vangeli, PUL 1998
C. Sigismondi, Gerbertus **9**, 91-94 (2016)

---

4  P. Couderc, *Les éclipses*, Presse Universitaire de France, Paris (1971),99
5  E. Fossat, et al. Astron. & Astrophys., *Asymptotic g modes: Evidence for a rapid rotation of the solar core*, **604**, A40 (2017).
6  J.-E. Solheim, *The sunspot cycle length – modulated by planets?*, Pattern Recogn. Phys., 1, 159–164, 2013
7  Eddington, A. S.; Plakidis, S., MNRAS **90**, p.65-71 (1929)